# Zombie Account Detection Based on Community Detection and Uneven Assignation PageRank

Qiu Yaowen[1], Li Yin[1], Lu Yanchang[1]
(1) BNU-HKBU United International College (UIC), 2000 Jintong Road, Zhuhai, China

**Abstracts**

In the social media, there are a large amount of potential zombie accounts which may has negative impact on the public opinion. In tradition, PageRank algorithm is used to detect zombie accounts. However, problems such as it requires a large RAM to store adjacent matrix or adjacent list and the value of importance may approximately to zero for large graph exist. To solve the first problem, since the structure of social media makes the graph divisible, we conducted a community detection algorithm – Louvain to decompose the whole graph into 1,002 subgraphs. The modularity of 0.58 shows the result is effective. To solve the second problem, we performed the uneven assignation PageRank algorithm to calculate the importance of node in each community. Then, a threshold is set to distinguish the zombie account and normal accounts. The result shows that about 20% accounts in the dataset are zombie accounts and they center in tier-one cities in China such as Beijing, Shanghai, and Guangzhou. In the future, a classification algorithm with semi-supervised learning can be used to detect zombie accounts.

## 1. Introduction

Zombie accounts refer to the fake fans on Weibo, which are usually malicious registered users automatically generated by the system. However, too many zombie accounts will make the graphical network between each account more complex, and each zombie account may only point to others, which seriously affects the quality of community classification. Therefore, before data analysis or classification, we need to screen out zombie accounts in advance to help us clean up the data set. We found that where there is a network, there will be out degrees and in degrees. So we can calculate the PageRank value and use PageRank algorithm. We can extend the PageRank algorithm to the social networks. We regard each account as a page and we should find out which account is less important. Because the PageRank of an account is larger, the account is more important and less likely to be a zombie account. This is the traditional PageRank social relation matrix. However, if an account is followed by some high PageRank accounts that it also should have high PageRank, so we need to use uneven assignation PageRank. However, there are some problems with PageRank. It requires huge RAM to store adjacent matrix/list and importance vector. It costs a long time to iterate the result. The behaviors of zombie fans make lots of divisible community. What's more, the importance of many nodes are approximate to zero (1e-8), unable to distinguish zombie account. As a result, we need to introduce community detection to optimize our approach. To complete the community detection task, we use Louvain algorithm as our algorithm to detect overall 1,002 community with acceptable modularity. Then, we applied the statistical method to consider accounts with extremely low importance value as detected zombie account. To measure the performance of the algorithm, we manually label 50 normal accounts and 50 zombie accounts and use our algorithm to scan the sample. The result suggests that the accuracy of the algorithm is 79%. The result also reveals that that around 20% of accounts in the dataset are zombie accounts and most of them located in tier-one cities in China. In the future, we expect to build a classifier with semi-supervised learning algorithm to classify each account or use the behaviors of users instead.

## 2. Related Work

### 2.1 PageRank Algorithm

PageRank (PR) is an algorithm used by Google Search to rank websites in their search engine results. PageRank was named after Larry Page, one of the founders of Google. PageRank is a way of measuring the importance of website pages. According to Google, PageRank works by counting the number and quality of links to a page to determine a rough estimate of how important the website is. The underlying assumption is that more important websites are likely to receive more links from other websites.

The PageRank algorithm outputs a probability distribution used to represent the likelihood that a person randomly clicking on links will arrive at any particular page. PageRank can be calculated for collections of documents of any size. It is assumed in several research papers that the distribution is evenly divided among all documents in the collection at the beginning of the computational process. The PageRank computations require several passes, called "iterations", through the collection to adjust approximate PageRank values to more closely reflect the theoretical true value.

### 2.2 Community Detection

In the real life and the Internet, there are numerous networks. Entities network likes urban transportation networks and power transmission networks, or virtual network likes social networks are all playing important roles. Among all sorts of different networks, social networks are the most representative and the most studied one since the scale is extremely large and it can reveal a large amount of social phenomena.

Herbert Simon proposed that the complex systems have modular structure characteristics in the first 1960s, which actually started the academic study of the community. The community reflects the local characteristics of individual behaviors in the network and their mutual relationships. In fact, community discovery has played a role in many fields: metabolic network analysis, gene regulatory network analysis, and master gene identification in the biological field such as predicting complexes and complexes in protein interaction networks. The community module is of great significance for understanding the organization and function of biological systems and the prediction of unknown protein functions. In addition, with the development of the internet and social applications, companies, governments, and scholars are also interested in studying the social networks for different purposes such as improving advertising recommendation system, fighting potential criminals, and performing social analysis.

In tradition, community refers to a group of nodes in the network that have a large similarity, resulting in a close internal connection and a sparse external group structure. First, using similarity to perform hierarchical clustering or spectral method were used in detecting the community. However, the average time complexity of this kind to method is approximately equal to $O(n^3)$ since it involves lots of calculation of matrix eigenvectors. Although some approximation methods at the expense of accuracy have been proposed later, in general, the efficiency of these algorithm still became a huge problem. Girvan and Newman [1] proved the existence of non-overlapping communities in many real networks and they also proposed the GN algorithm, which started the research on community detection. Later, Newman [2] proposed a widely used community structure measurement function called Modularity, turned the community detection problem into an optimization problem and unparalleled the development of non-overlapping community detection research.

## 3. Methodology
### 3.1 Uneven Assignation PageRank

PageRank is calculated by the page's hyperlink to evaluate the pages in the view of quantity and quality. However, we can apply this method into the community to detect the zombie account.

Let's see how to calculate PageRank first. There are two concepts: in-degree and out-degree. The in-degree means the number of links that come in and out-degree means the number of links that get out. The weight of each page is calculated by this function:

$$P(u) = \sum_{v \in B_u} \frac{P(v)}{L(v)} \quad (1)$$

The u means the pages needed to evaluate. $B_u$ is the set of in-degree links. Through this function, we can calculate each page's weight in the set and get a PageRank matrix.

However, we regard each account as a page and we should find out which account is less important because the PageRank of an account is larger, the account is more important and less likely to be a zombie account. In real life, when an account concerns to the other N accounts, this account needs to evenly assign its PageRank value to the other N accounts. However, the weights of the N concerned accounts are different, so this even assignation way will affect the order of the actual PageRank. If the account is concerned by more fans, then this account's PageRank will be higher and if an account is followed by some high PageRank accounts that it also should have high PageRank. Then here we use a new work called IO to represent the credibility of the account according to its fans and follow accounts.

$$IO = \frac{FanNum}{FollowNum + FanNum} \quad (2)$$

Fan-Num is the number of in degree of the account, namely the fans amount, while the FollowNum is the number of out degrees. Then we should calculate each account's IO value and they will be needed to find the new social matrix. Each account's value in the social relations matrix is calculate by this function:

$$K_v = \frac{IO_v}{\sum_{i=1}^{n} IO_i} \quad (3)$$

The new value in the matrix will be based on the IO value of each account, this upper one is the account the target account's IO value and the bottom one is the sum of all output account's IO value. Through this uneven assignation PageRank method, we can easily find out the zombie accounts with very low PageRank in the communities.

### 3. 2 Modularity

To measure the performance of community detection algorithm, Commonly, the modularity Q is used:

$$Q = \frac{1}{2m} \sum_{i \neq j} \left( A_{ij} - \frac{k_i k_j}{2m} \right) \delta(C_i, C_j) \quad (4)$$

where $A_{ij}$ is an element in the adjacent matrix in the network (if node i is connected with node j, then A = 1 else 0). $C_i$ represents the community that node i located in. m is the total number of edges in the network. $\delta(C_i, C_j) = 1$ if node i and j locate in the same community, else $\delta(C_i, C_j) = 0$. $k_i$ represents the degree of node. $k_i = \sum_w A_{iw}$.

Modularity is a measurement method to evaluate the division of a community network. Its physical meaning is the difference between the sum of the weights of the connected edges of nodes in the community and the sum of the weights of the connected edges under random conditions. It ranges from [-0.5, 1].

The Q value is calculated during each iteration, and the maximum value of Q is the ideal division of the network. The Q value ranges from 0-1. The larger the Q value, the higher the accuracy of the community structure divided by the network. In actual network analysis, the highest point of the Q value generally appears between 0.3-0.7.

### 3.3 Community Detection Algorithms

After defining the measurement, we will implement some algorithm for nonoverlapping community detection and overlapping detection. For non-overlapping community detection, we plan to use algorithms based on modularity optimization such as Fast unfolding (Louvain) [4] and dynamic algorithms such as Infomap [7]. We compare the result from each algorithm and choose the result with the largest Q as the final result. And we will also compare the time consuming. In addition, detecting overlapped community is also important since it really exists in lots of real scenarios. We adopted the CPM algorithm and get the result of overlapping community.

The Louvain algorithm mainly has two stages: First, Each point sets as a community. Then consider the neighbor nodes of each community and merge into the community. Find the largest positive modularity Q and merge the point to community. It could be better to repeat this stage many times until the modularity Q without changing. Second, the second phase of the algorithm consists in building a new network whose nodes are now the communities found during the first phase. To do so, the weights of the links between the new nodes are given by the sum of the weight of the links between nodes in the corresponding two communities. Links between nodes of the same community lead to self-loops for this community in the new network.

As Table 1 illustrates the pseudocode of Louvain algorithm. Typically, i represents a node in the graph, j is another node (except node i itself) in the graph, $and\ A_{ij}$ is the weight between node i and node j, it is assigned the value 1 if node i and j are adjacent else it is assigned the value 0. Character m is the sum of all the weight, $k_{i,in}$ is the sum of edge weights of all nodes in the community of node I and $K_i$ is the sum of edge weights of all nodes in the whole graph of node i.

*Table 1 Pseudocode of Louvain algorithm*

| Algorithm 1. Louvain |
|---|
| **Input**: i, A |
| **Output**: community |
| **initialize** each node i as a community |
| calculate $A_{ij}$ |
| calculate m |
| **repeat** |
|     **repeat** |
|         calculate each i's Q with j |
|         calculate each i's △Q with j |
|         merge node i into the largest Q's community |
|         until no change |
|     **repeat** |

```
        update the $A_{ij}$
        update the m
        update the $k_{i,in}$
        update the $K_i$
      until stopping criterion is met
   until reaches the maximum iteration times
   return Louvain community information
```

### 3.4 Dataset

We choose Sina Weibo, the most influential social media in China, as our research object. We will use the dataset provided in a paper [3] which studies the social influence locality for modelling retweeting behaviors. The data set was crawled in the following ways: To begin with, 100 random users were selected as seed users, and then they were collected. The crawling process produced in total 1.7 million users and 0.4 billion following relationships among them, with average 200 followees per user. For each user, the crawler collected her 1,000 most recent microblogs (including tweets and retweets). The process resulted in totally 1 billion microblogs. The dataset also includes all the users' profiles with contain name, gender, verification status, and microblogs. Table 1 shows the statistic of the dataset.

*Table 2 Statistics of the dataset*

| Dataset | Users | Relationships | Microblogs | Retweets |
|---|---|---|---|---|
| Weibo | 1,776,950 | 308,489,739 | 300,000 | 23,755,810 |

The dataset includes 4 parts: *uidlist, user_profile1, user_profile2, weibo_network*. First, *uidlist* provides the original id of each Weibo user in the dataset. *User_profile1* and *user_profile2* both include all the information of each user which not only has the basic information like name, city, but also give message such as the number of the reciprocal relationships, the number of followers, the number of published/retweeted tweets and so on. Main dataset is the *weibo_network*. This file describes the Sina Weibo sub-network at the very first timestamp. First line consists of two integers, representing the number of users N and number of the follow relationships M respectively. In the following N lines, each line starts with an integer v1_id, representing the *user id* of user v1, followed by another integer k. And the following 2k numbers describes the user followed by v1, each represented by a user id v2_id and a number indicating the type of relationship. In *weibo* data set, "1" indicates that user v1 and v2 have a reciprocal follow relationship, while "0" indicates that their relationship is not reciprocal. After reorganizing the dataset, we figure out 116,815,889 edges in total.

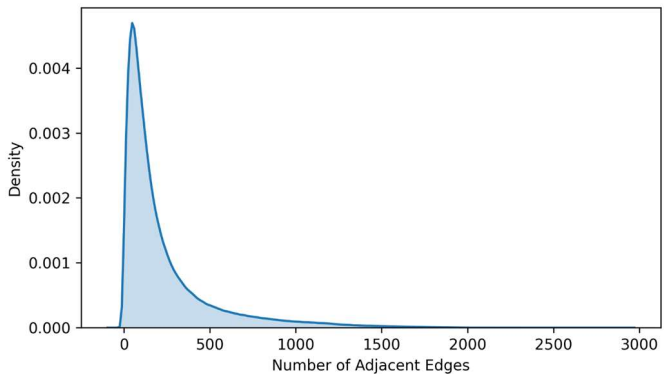

*Figure 1 Kernel Density Estimation of the distribution of numbers of Adjacent Edges*

As Figure 1 shows the kernel density estimation of the distribution of numbers of adjacent edges of all nodes. The distribution is seriously right skewed which means most of the amount of adjacent edges are less than 500. And some of them exceeds 1500. The total distribution is imbalance and will inevitably increase the time complexity of algorithms.

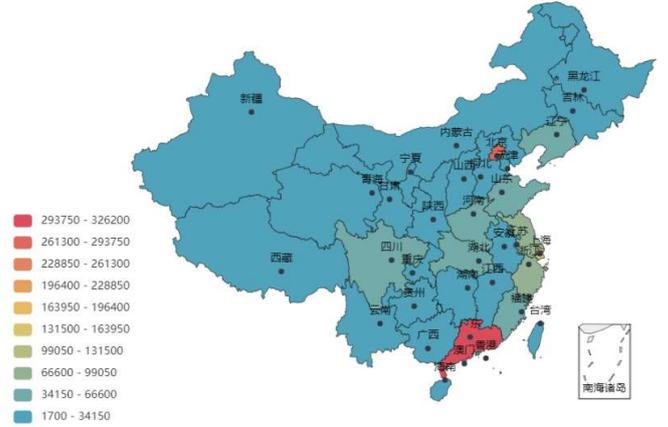

*Figure 2 The region distribution of the users*

Figure 2 illustrates the region distribution of users in the dataset. The distribution is polarized since most of the users locate in Guangdong province and Beijing. Except these users, most of the rest are in coastal areas and Sichuan province. The proportion of users located in northwest, northeast, and southwest is low.

### 3.5 Detect Zombie Account using outliers

Instead of detecting zombie account in the whole dataset, we detected it in each separate community. After executing the uneven assignation PageRank algorithm, the importance vector of all nodes in each community is available and the summation of the value should be all equal to 1. It's intuitive that considering those nodes with extremely low PageRank value as zombie account. Since it is one-dimensional data, it is reasonable to set a threshold that any node with PageRank value lower than this threshold should be considered as zombie account. It is defined as:

$$IQR = Q3 - Q1 \qquad (5)$$
$$threshold = Q1 - 1.5 * IQR \qquad (6)$$

Q1 and Q3 represent the $25_{th}$ percentile and $75_{th}$ percentile of the vector of PageRank value in each community. Thus, IQR is the difference between Q3 and Q1. The threshold is set to be equal to the difference between the $25_{th}$ percentile and 1.5 IQR. If an account's PageRank value is lower than the threshold, it should be considered as zombie account.

As Figure 3 demonstrates the boxplot of a community without zombie account and Figure 4 demonstrates the boxplot of a community with several zombie accounts. It is apparently that there is no node with PageRank value lower than the threshold in Figure 3 and there are several nodes in red in the left side that lower than the threshold in Figure 4. These nodes are considered as zombie account.

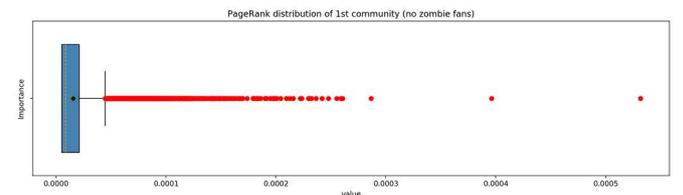

*Figure 3 Boxplot of community without zombie account*

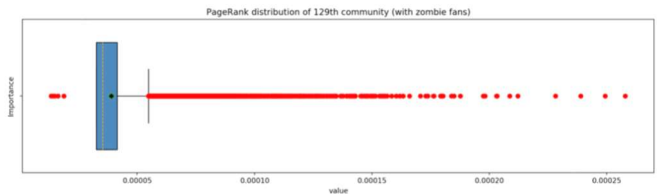

*Figure 4 Boxplot of community with zombie account*

## 4 Result
### 4.1 Community Detection

Table 3 illustrates the result of community detection by Louvain algorithm. The number and edge of the node is the same as the dataset shows. The algorithm divides 1,002 community on the whole dataset with modularity equals to 0.5658. According to the definition of modularity, the quality of community detection is considered as effective.

*Table 3 Result of community detection by Louvain algorithm*

| Type | Nums |
| --- | --- |
| Nodes | 1,776,950 |
| Edges | 58,409,237 |
| # Community | 1,002 |
| Modularity | 0.5658 |

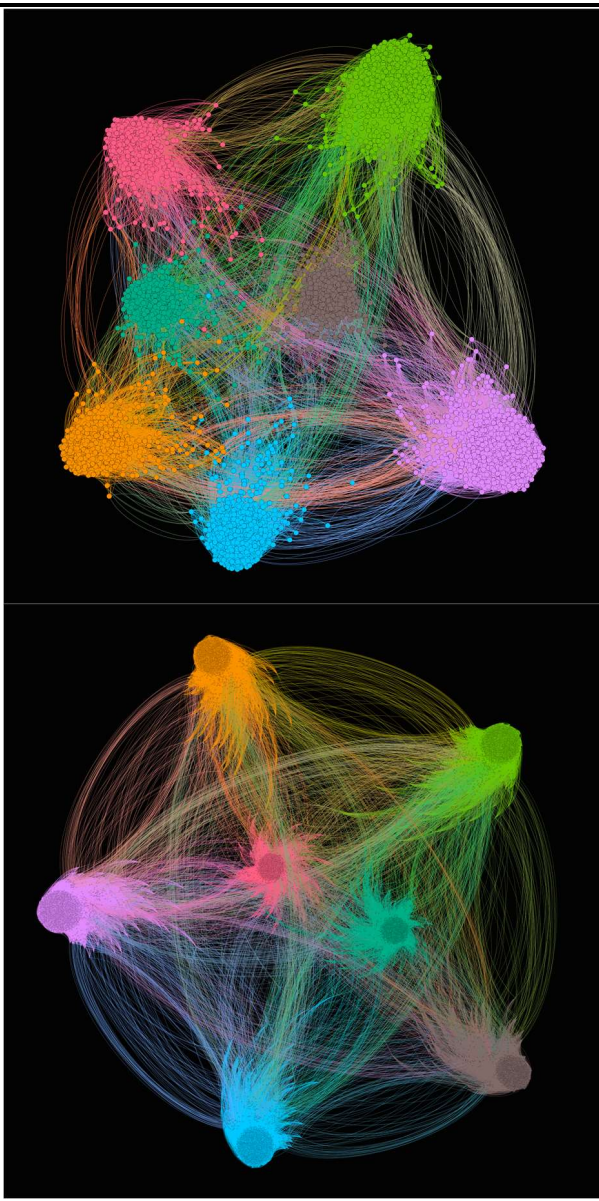

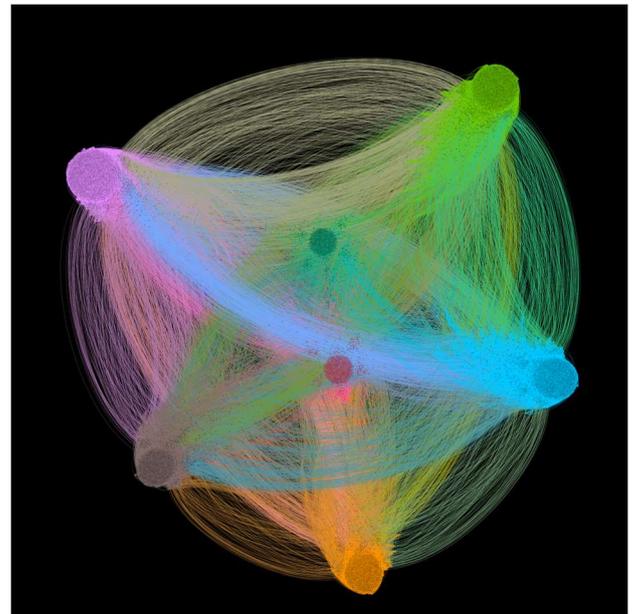

*Figure 5 Visualization of community detection of small size, medium size, and large size. From top to the bottom,*

Figure 5 illustrates the Visualization of community detection of small size (with number of nodes between 500 to 1000), medium size (with number of nodes between 1000 to 2000), and large size (with node larger than 5,000), from left to the right, respectively. The visualization shows that the community detection algorithm can effectively clusters community with different scale. And it also shows that the connections across community become more and more significant when the size of community increases. It may suggest that the detection algorithm may work well on the small community but work slightly worse in the larger community.

### 4.2 Zombie Account Detection

After executing the PageRank algorithm on all the community, the classification of zombie account is available. In the whole dataset, there are over 190,000 zombie accounts are detected by the algorithm and the proportion of zombie account in the whole dataset is 19.6%, as the Figure 6 shows. Figure 7 illustrates the location of zombie account based on the dataset, it is apparently that most of the zombie accounts are located in one-tier cities in China such as Beijing, Shanghai, and Guangzhou. The most likely reason is that zombie accounts are used by some companies that spreading public opinion. Since the dataset was collected in 2014, the proportion of zombie accounts is more likely to increase nowadays. And with the development of the internet, many bloggers eager to increase their 'visual' fans number or some companies need public opinion to increase their sales. These jobs are done by most of media company located in one-tier cities, thus the proportion of the location of zombie fans in one-tier city is likely to increase either.

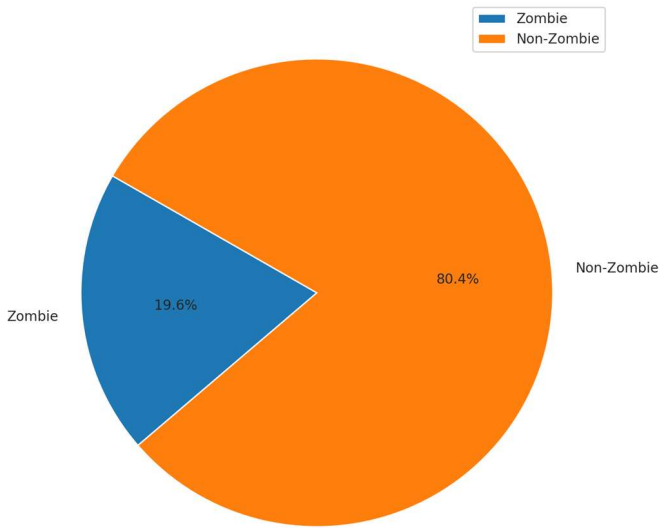

*Figure 6 Proportion of zombie accounts in the dataset*

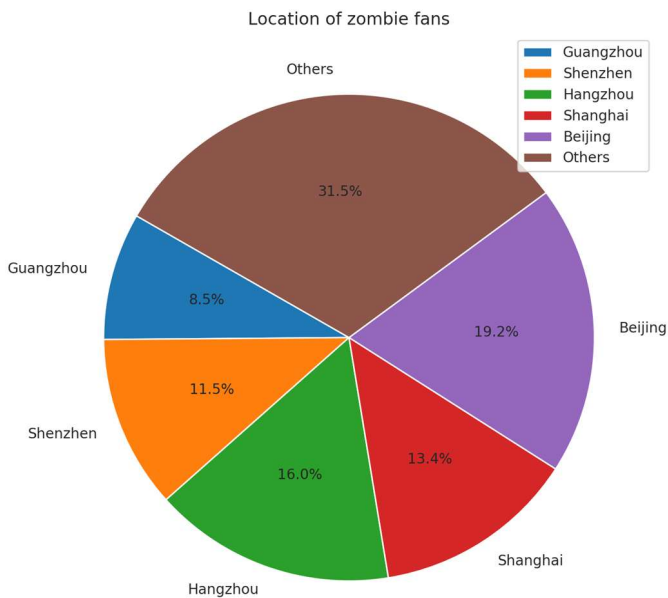

*Figure 6 Location of all zombie account*

**4.3 Performance of the algorithm**

It is reasonable to combine the community detection algorithm and uneven assignment PageRank algorithm to detect zombie account. However, misclassification problem always exists. It is time-consuming to verify the whether the classification on each account is correct or wrong. In order to measure the performance of our model, we randomly select 50 zombie account and 50 non-zombie account in the dataset and use the result of zombie account detection to draw a confusion matrix, as Table 4 demonstrates.

The result shows that most of the classification is correct and take most of the proportion. And the number of classifying Non-Zombie Account to a Zombie Account is two times the number of classifying Zombie-Account to Non-Zombie Account. It means the algorithm tends to classify an account to a zombie Account.

*Table 4 Confusion matrix of 100 manual labeled sample*

|  | Zombie Account | Non-Zombie Account |
|---|---|---|
| Zombie Account | 41 | 17 |
| Non-Zombie Account | 9 | 33 |

Figure 8 is the numerical performance of the algorithm. Overall, the accuracy of the algorithm is around 74%. The score of recall is 81% and the precision score is 89%. The general score, f1-score is around 83%. The data also suggest that the algorithm prefers to classify a normal account into zombie account. Thus, the threshold can be set to lower value or we should set more constraints of the algorithm to prevent it classify a normal account to a zombie account.

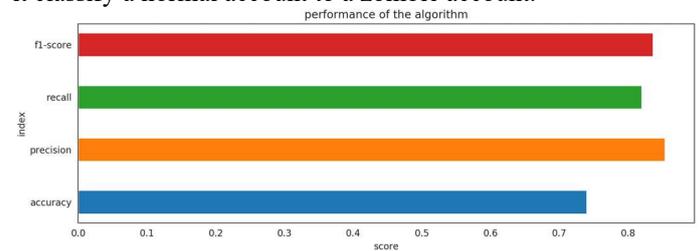

*Figure 7 Performance of the algorithm*

**5   Future Works**

There are possible improvements for our projects. First, a classifier can help to classify zombie account. In the original dataset, there are many features related to a typical account: number of followers and followees, register date, region, post, etc. A classical machine learning algorithm such as support vector machine and decision tree can be used to complete this binary classification task. Since it is impossible to label all the data in the dataset, it is wise to use semi-supervised learning algorithm to do this job. In the second, instead of using the relationships between users, users' behaviour can be used to detect zombie account. Paper *Optimizing Opinions with Stubborn Agents Under Time-Varying Dynamics* by D. Scott Hunter, Tauhid Zaman from MIT can be referred

**6   Conclusion**

Zombie accounts refer to the fake fans on Weibo, which are usually malicious registered users automatically generated by the system. It is not conducive to data analysis and classification. PageRank algorithm is an algorithm able to compute the importance of nodes in the graph network. However, it exists some problems in our project. First, it requires the computer to store the adjacent matrix or adjacent list in the RAM. It is hard for our machine to store such a big data in the computer. Second, for a graph with millions of even billions of nodes, the importance of each node is quite small or approximate to zero. It is difficult to tell the difference of importance of nodes with similar value. To solve the first problem, considering the links between nodes make graph divisible while keeping the main structure of it, a community detection algorithm – Louvain is used to divide the graph into several communities. To solve the second problem, the uneven assignment PageRank algorithm is used. It does not ignore the contribution of each node: node with more fans number is more important and contribute more than the other node. For each importance vector in the community, a threshold is set to distinguish zombie account and normal account.

A weibo social dataset contains 1.3 million nodes and 50 million edges is used. The result suggests that there are total 1,002 communities with modularity 0.58. The uneven PageRank algorithm detect about 18 thousand zombie accounts, which take the proportion of about 20% of the dataset. The location of this zombie account are centred in one-tier cities in China, such as Beijing, Shanghai, and Guangzhou.

In the future, a classification algorithm with semi-supervised learning can be helped to classify zombie accounts, take features of an account like register date or number of fans. Or user's behaviours instead of the relationship in the graph can be used to do so.